\documentclass[acmlarge,menuscript,sigconf]{acmart}
\AtBeginDocument{%
  }

\copyrightyear{2026}
\acmYear{2026}
\setcopyright{cc}
\setcctype{by}
\acmConference[LAK 2026]{LAK26: 16th International Learning Analytics and Knowledge Conference}{April 27-May 01, 2026}{Bergen, Norway}
\acmBooktitle{LAK26: 16th International Learning Analytics and Knowledge Conference (LAK 2026), April 27-May 01, 2026, Bergen, Norway}
\acmPrice{}
\acmDOI{10.1145/3785022.3785035}
\acmISBN{979-8-4007-2066-6/2026/04}

\begin{document}

\title{AI Knows Best? The Paradox of Expertise, AI-Reliance, and Performance in Educational Tutoring Decision-Making Tasks}


\author{Eason Chen}
\affiliation{%
  \institution{Carnegie Mellon University}
  \city{Pittsburgh}
  \state{PA}
  \country{USA}
}

\author{Jeffrey Li}
\affiliation{%
  \institution{Carnegie Mellon University}
  \city{Pittsburgh}
  \state{PA}
  \country{USA}
}

\author{Scarlett Huang}
\affiliation{%
  \institution{Carnegie Mellon University}
  \city{Pittsburgh}
  \state{PA}
  \country{USA}
}

\author{Xinyi Tang}
\affiliation{%
  \institution{Carnegie Mellon University}
  \city{Pittsburgh}
  \state{PA}
  \country{USA}
}

\author{Jionghao Lin}
\authornote{Corresponding Author}
\affiliation{%
  \institution{The University of Hong Kong}
  \city{Hong Kong}
  \country{China}
}


\author{Paulo F. Carvalho}
\affiliation{%
  \institution{Carnegie Mellon University}
  \city{Pittsburgh}
  \state{PA}
  \country{USA}
}

\author{Kenneth R. Koedinger}
\affiliation{%
  \institution{Carnegie Mellon University}
  \city{Pittsburgh}
  \state{PA}
  \country{USA}
}

\renewcommand{\shortauthors}{Chen et al.}
\makeatletter
\let\@authorsaddresses\@empty
\makeatother



\begin{abstract}
We present an empirical study examining how experienced tutors (experts) and non-tutors (novices) evaluate the correctness of tutor praise responses under different AI-assisted decision-support interfaces and explanation styles. We examine human-AI reliance patterns by decomposing interaction errors into \emph{over-reliance} (accepting incorrect AI suggestions) and \emph{under-reliance} (rejecting correct AI suggestions), together with time cost as a process-level indicator.
Across conditions, human-AI collaboration improved accuracy compared to humans working alone, but consistently underperformed an AI-only baseline, indicating that human judgment introduced additional errors even when assisted by a highly accurate model. Novices benefited more from AI support since they tend to follow AI suggestions, whereas experts frequently overrode correct AI advice, resulting in lower overall performance, revealing a paradox of expertise in educational decision-making.
We further compare two explanation modalities: textual reasoning and inline highlighting. Textual reasoning reduced under-reliance when the AI was correct but increased over-reliance when the AI was wrong, while inline highlighting exerted minimal influence on either behavior. Notably, neither explanation modality improved accuracy, and both increased time costs.
As a contribution to learning analytics, we demonstrate how reliance patterns (over-reliance and under-reliance) and time cost function as process-level indicators that reveal how users integrate, or fail to integrate, AI recommendations. Our findings underscore the need for adaptive, trust-calibrated explanation strategies in tutor-facing decision support systems that balance accuracy, efficiency, and accountability in human-AI collaboration.
\end{abstract}

\ccsdesc[500]{Human-centered computing~Empirical studies in HCI}
\ccsdesc[500]{Human-centered computing~User interface design}

\keywords{Human-AI Collaboration, Human-AI Interaction, Educational Decision Making, Explainable AI, Tutor-Facing Decision Support}


\maketitle

\section{Introduction}
Many educational tasks, such as grading~\cite{armfield2025avalon,chen2025generative}, assessment~\cite{anderson2003classroom,han2024improving}, and providing feedback~\cite{thomas2023tutor,hattie2007power,lin2023using, Lin2024,zhao2025slideitright}, require educators to make critical judgments about student performance and progress. These tasks are inherently \emph{decision-making} processes: teachers assess student outputs against learning objectives, deciding how best to score, comment on, or otherwise respond to student work. As educational settings become more complex and class sizes grow, educators often face increasing time and cognitive demands to maintain fair and consistent evaluations \cite{jerrim2020teacher,kreuzfeld2022teachers}. When educational assessments are deliberately designed and decomposed into analytic criteria, scoring often reduces to a sequence of binary evidence checks (e.g., present/absent or meets/does not meet) that are subsequently aggregated into partial-credit scores or performance levels. This evidence-centered operationalization is common in learning analytics and teacher-facing decision-support systems, creating clear opportunities for AI to assist in analyzing data and help close the loop from analytics to decision 
\cite{li2024bringing,cheng2024evidence,clow2012learning,dourado2021teacher,fernandez2024data}.

In the present era, AI significantly aids human decision-making. Studies show that AI recommendations can improve speed and performance as well as introduce reliance in tasks such as question-answering~\cite{vasconcelos2023explanations}, logical reasoning~\cite{bansal2021does,si2023large}, sentiment analysis~\cite{bansal2021does}, and image annotation~\cite{morrison2023evaluating}. Nevertheless, prior work predominantly employed traditional machine-learning methods for non-educational tasks. Little research has examined how the latest Large Language Models (LLMs) could support decision-making in educational contexts, specifically in tutoring. Unlike these tasks, tutoring requires accountability and specialized pedagogical knowledge (e.g., how to give constructive feedback). Given the context-dependent nature of AI’s impact, applying findings from non-educational domains directly to tutoring settings with LLMs poses inherent risks.


These circumstances create a strong incentive to incorporate AI tools that reduce educators' cognitive load and help them make effective decisions or feedback~\cite{lin2023using}. For instance, AI-driven feedback could help educators quickly assess how well they are motivating or guiding students. Yet, as discussed above, research on human-AI decision-making in education, especially in tutoring scenarios, remains limited. This paper aims to bridge this gap by examining how LLMs can facilitate effective decision-making in evaluating tutor responses, determining whether certain praise or feedback is suitable. Specifically, we address the following \textbf{R}esearch \textbf{Q}uestions:

\noindent \textbf{RQ1}: In evaluating the appropriateness of tutor-praise responses, how does performance compare between (a) humans working alone and (b) human–AI collaboration?

\noindent \textbf{RQ2}: How do different tutor experience levels and AI explanation styles (textual reasoning vs. inline highlighting) influence human performance, specifically with respect to accuracy, time efficiency, and reliance on AI?

In response to those research questions, we developed two explanation interfaces and observed that textual reasoning fosters greater reliance on AI; further, humans spend more time when collaborating with AI. Finally, while novices boost their overall accuracy by trusting the AI’s generally correct suggestions (yet over-rely when it is wrong), experienced tutors often override correct AI advice, resulting in under-reliance.


\section{Related Work}

\subsection{AI-Assisted Decision Making}

Prior research on AI-assisted decision-making shows mixed results: humans may be misled by AI errors, especially with unclear or complex explanations \cite{vasconcelos2023explanations,buccinca2021trust,dietvorst2015algorithm}, yet well-designed explanations can help users detect unreliability and avoid blind reliance \cite{morrison2023evaluating}. Visual errors often mislead more than textual ones, and conflicting cues may prompt users to question AI \cite{morrison2023evaluating}. Overall, humans tend to either \emph{over-rely} on wrong AI suggestions (accepting errors uncritically) or \emph{under-rely} on correct AI suggestions (ignoring accurate advice), and thus rarely achieve \emph{appropriate reliance}, where users trust AI when it is right and reject it when it is wrong \cite{vereschak2021evaluate}. Moreover, research show that different explanation designs such as inline highlights can reduce cognitive load \cite{bansal2021does}. Reliance also varies by expertise: non-experts are more susceptible to over-reliance \cite{morrison2023impact}.
In this study, we examine how different explanation styles affect human–AI collaboration in evaluating tutoring feedback for users with different levels of expertise, and discuss their implications for learning analytics in educational contexts.

\subsection{Giving Effective Praise in Tutoring and Evaluating it with AI}

Prior work in tutor training emphasizes \emph{effective praise} as feedback that is specific to what the learner did, focusing on process or effort rather than personal traits or outcomes \cite{thomas2023tutor}. For example, \textit{``You checked each step carefully and revised when something didn’t look right, keep using that approach''} praises the learner’s strategy, whereas \textit{``You’re so smart!''} or \textit{``Perfect, you got it right''} focuses on inherent ability or just the outcome. This distinction aligns with growth-mindset research, which shows that attributing success to controllable strategies and effort encourages persistence and continued engagement, while ability-focused praise can promote fixed beliefs \cite{blackwell2007implicit}.
Building on these guidelines, prior work has employed AI to classify tutor responses, such as using a BERT-based Named Entity Recognition (NER) model to classify effective praise~\cite{lin2023using}, obtaining up to 73\% accuracy and a F1 score of 0.81. They found that the model performed well for effort-related praise but struggled with outcomes-based praise due to the lack of training data.


\section{Method}

\begin{figure*}[h]
    \centering
    \includegraphics[width=1\linewidth]{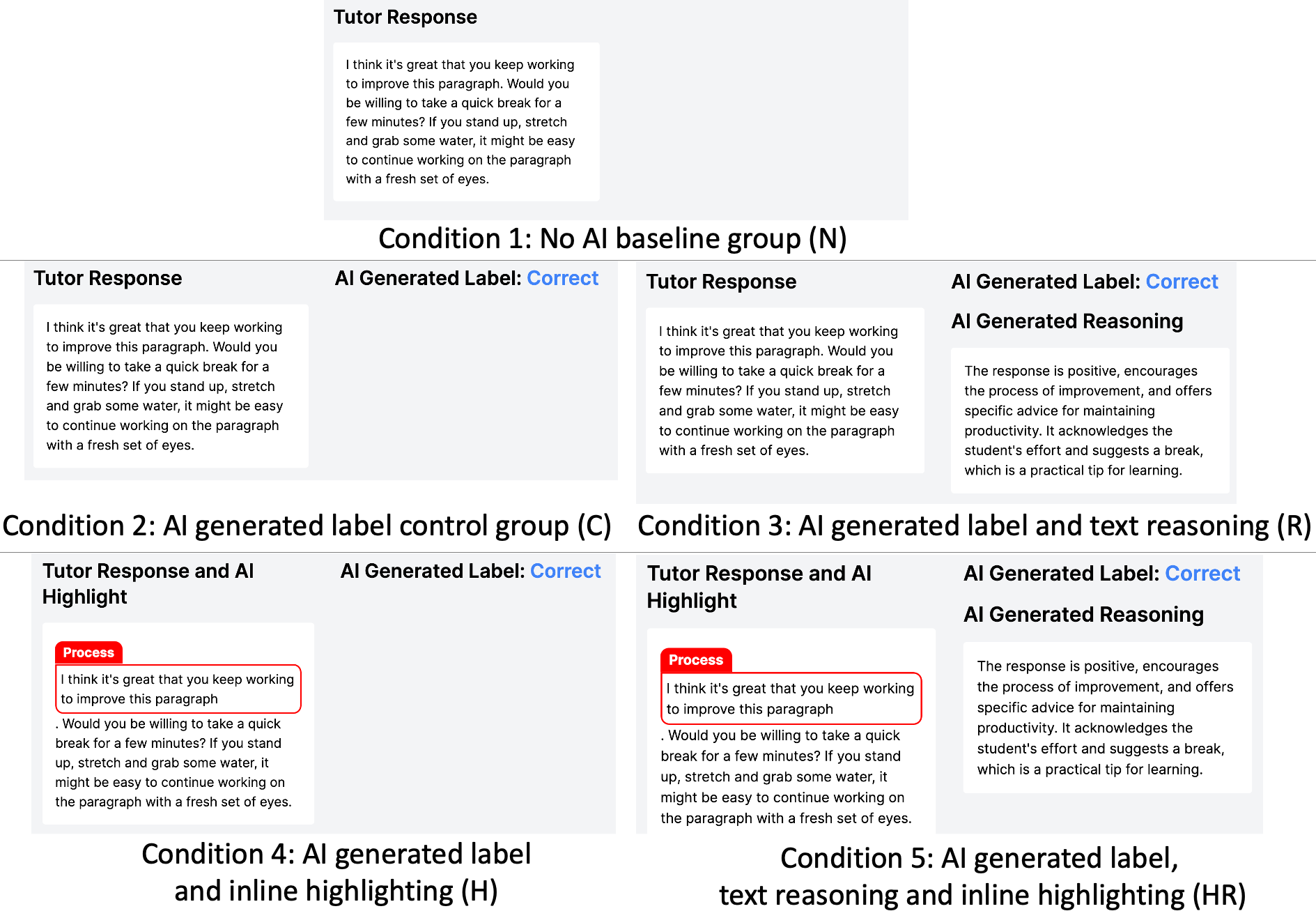}
    \caption{The five different interfaces we designed for AI-assisted decision-making.}
    \label{fig:AI_BIAS_Conditions}
\end{figure*}

\subsection{Data Preparation}
Through a tutor training system, we collected 216 responses from a simulated scenario where tutors complimented students after solving a math problem. Two senior and experienced tutors and a PhD learning scientist were invited to label the correctness of these responses. The kappa score between the two experienced tutors was 0.76, indicating substantial consistency, and then the learning scientist resolved the disagreements. This process established our ground truth data. Of the total responses, 155 were labeled as correct (effective praise), and 61 were labeled as incorrect (not effective praise). Among these, four responses were used as a few-shot training data in the prompt, while the rest were used for prediction.

\subsection{Prompt Engineering and Model Selection}
We prompted an LLM to assign a binary label and produce the following two explanations: text reasoning and inline highlights (by HTML tags) over the original input. Based on our prior evaluation on this dataset \cite{chen2025identifying}, \texttt{gpt-4-0613} performed best with 0.88 accuracy and 0.92 F1 score, so we use its outputs in the user study. For more about prompt engineering and evaluation on this dataset, please refer to our prior work at \cite{chen2025identifying}.


\subsection{Interface Design} \label{interface-design}

In order to investigate user reliance on AI in different explanation methods, we designed five different interfaces as shown in \autoref{fig:AI_BIAS_Conditions}. The AI-generated label, Reasoning, and Highlightings are from \textit{gpt-4-0613}. These five interfaces are:

\begin{enumerate}
    \item N: Task content without AI (\textbf{N}o AI baseline group).
    \item C: Task content with AI-generated label (\textbf{C}ontrol group with AI)
    \item R: Task content with AI-generated label and text \textbf{R}easoning.
    \item H: Task content with AI-generated labels and inline \textbf{H}ighlightings.
    \item HR: Task content with AI-generated labels, \textbf{H}ighlightings, and \textbf{R}easoning.
\end{enumerate}

We use the No AI group as a \textbf{baseline} to compare users' performance with and without AI. Moreover, we use the AI label-only control group to compare users' performance and reliance on different types of AI explanations.

\subsection{Participants}

We recruited a total of 95 participants by convenience sampling, including 47 professionally trained tutors and 48 participants from Prolific. The age distribution varied between the two groups: Prolific participants ranged from 23 to 73 years old, with a mean age of 39.2 years (SD = 14.5), while tutors were between 18 and 29 years old, with a mean age of 22.3 years (SD = 2.6). In terms of gender distribution, the Prolific group consisted of 28 male and 20 female participants, whereas the tutor group comprised 17 male and 30 female participants.

\subsection{Procedure and Design}

Our experiment was approved by the Institutional Review Board (IRB). At the experiment platform, participants started by going through a training module on giving effective praise, which consisted of guidelines and examples explaining why certain tutors' responses are or are not effective praise. They then proceeded to engage in a task involving five rounds of judging the correctness of Tutor Responses. We employed a counterbalanced design, in which each participant was randomly assigned to one of the possible permutations of the five conditions in \autoref{interface-design}, ensuring that each condition occurred in each position equally often across participants. Each condition included 20 tasks.

Moreover, the tasks encountered by the participants were also randomized, but we kept these tasks' distribution and accuracy rates aligned with the original data. 
That is, 
in the 20 tasks of each design condition, participants encountered 13 correct tutor responses and 7 incorrect tutor responses. Moreover, 
to reflect the distribution of real data, we set the AI's accuracy rate to 0.8, and participants encountered 3 false negatives and 1 false positive at each design condition. 
Finally, to prevent participants from losing confidence in the AI due to errors, we ensured that participants received correct suggestions from the AI for the first five tasks across all conditions.

\begin{table*}[h]
    \centering
    \caption{Decision-Making Performance Comparison Between Tutor and Prolific Participants. T-test Results. }
    \begin{tabular}{c c c c l c}
        \hline
        Condition & Tutor Mean (SD) & Prolific Mean (SD) & t-statistic & p-value & Effect Size \\
        \hline
        No Label & 0.696 (0.133) & 0.714 (0.107) & -0.757 & 0.4512 & 0.154 \\
        AI Label Only & 0.703 (0.132) & 0.742 (0.091) & -1.676 & 0.0970\textsuperscript{+} & 0.342 \\
        Text Reasoning & 0.726 (0.122) & 0.748 (0.109) & -0.949 & 0.3450 & 0.194 \\
        Inline Highlighting & 0.707 (0.108) & 0.762 (0.081) & -2.818 & 0.0059** & 0.575 \\
        Inline Highlighting + Text Reasoning & 0.729 (0.105) & 0.768 (0.076) & -2.129 & 0.0359* & 0.435 \\
        Overall Average & 0.712 (0.088) & 0.747 (0.057) & -2.302 & 0.0236* & 0.470 \\
        \hline
        \\[-1.8ex]
        \multicolumn{6}{l}{$\textsuperscript{*}p < 0.05$, $\textsuperscript{**}p < 0.01$, $\textsuperscript{***}p < 0.001$}
    \end{tabular}
    \label{tab:performance_result}
\end{table*}

\begin{figure*}[h]
    \centering    \includegraphics[width=0.9\linewidth]{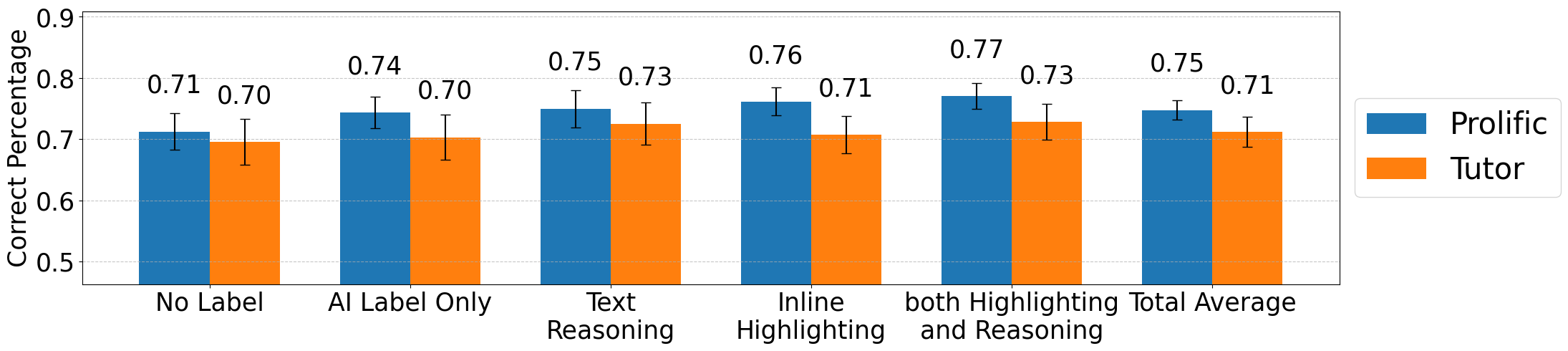}
    \caption{Correct Percentage across different conditions and groups. Participants from Prolific and working with AI performed better. The error bar is Standard Error.}
    \label{fig:correct_percentage}
\end{figure*}

\subsection{Data Collection and Analysis}

We first conducted a 2 x 2 Analysis of Variance (ANOVA) to compare how different groups of participants perform with and without AI label and explanation. 
Moreover, we used a one-way ANOVA with post hoc tests to compare error rates on AI-correct and AI-incorrect items between the No Label baseline and AI-assisted conditions.
Then, we used repeated measures ANOVA with the presence of Highlighting and Text Reasoning in a 2 x 2 ANOVA to compare participants' performance and reliance on different types of explanations while the label is always presented. Finally, we compared the difference between the experienced Tutor and Prolific participants using an independent sample t-test for the following measures:

\subsubsection{Human Performance}
We evaluated the participants' performance based on their accuracy rate when they performed data labeling decision-making tasks in a specific condition.

\subsubsection{Time Spent}
We evaluated the average time participants spent on each task by calculating the time interval between their task submissions. When analyzing the time spent, we noted that since the experiment was conducted online, a few participants disconnected midway, engaged in other activities, and then reconnected later, resulting in some extreme completion times. To address this, we applied the \textbf{Interquartile Range (IQR) method to remove outliers} before conducting our analysis, as recommended by \cite{vinutha2018detection}.

\subsubsection{Reliance on AI in Different Conditions}

Building on prior research \cite{vasconcelos2023explanations,buccinca2021trust,dietvorst2015algorithm,morrison2023evaluating}, we categorize two forms of reliance on AI as follows, according to how participants perform in testing scenarios where the AI provides labels and explanations:

\begin{itemize}
    \item \textbf{Over-Reliance}: When the AI provides an incorrect suggestion, and the user follows the suggestion, resulting in an incorrect decision. 
    \item \textbf{Under-Reliance}: When the AI provides a correct suggestion, but the user ignores it and makes an incorrect decision.
\end{itemize}

\section{Results}






\subsection{Decision-Making Performance}


The results comparing the performance of Tutors and Prolific participants across different task conditions are presented in \autoref{fig:correct_percentage} and \autoref{tab:performance_result}.
We first conducted a 2-way ANOVA to compare how different groups of participants perform with and without AI. The analysis revealed a \textbf{significant main effect of AI assistance} (\( F(1, 471) = 6.83, p = .009, \eta_p^2 = .014 \)), indicating that participants using AI performed significantly better than those without AI. Moreover, there was a \textbf{significant main effect of group} (\( F(1, 471) = 5.34, p = .021, \eta_p^2 = .011 \)), indicating that participants from Prolific have better overall performance compared to experienced tutors. The interaction effect between AI assistance and group was not significant (\( F(1, 471) = 0.91, p = .341, \eta_p^2 = .002 \)). 

We then conducted a comparison of repeated measures in terms of the presence of \textit{Text Reasoning} and \textit{Inline Highlighting} while the AI label is always present. The results showed no significant main effects or interactions. 






\begin{table*}[h]
    \centering
    \caption{Comparison of average time spent in seconds after IQR filtering across conditions for Tutor and Prolific participants, including in-significant t-test results.}
    \begin{tabular}{c c c c c c}
        \hline
        Condition & Tutor Mean (SD) & Prolific Mean (SD) & t-statistic & p-value & Cohen's d \\
        \hline
        No Label & 6.65 (3.95) & 7.24 (3.43) & -0.600 & 0.551 & -0.160 \\
        AI Label Only & 5.88 (3.36) & 6.87 (3.30) & -1.122 & 0.267 & -0.297 \\
        Text Reasoning & 10.34 (6.40) & 11.88 (5.13) & -0.995 & 0.324 & -0.266 \\
        Highlight & 11.11 (7.58) & 12.56 (7.61) & -0.722 & 0.473 & -0.191 \\
        Highlight + Text Reasoning & 11.86 (7.12) & 12.10 (6.04) & -0.140 & 0.889 & -0.037 \\
        \hline
    \end{tabular}
    \label{tab:avg_time}
\end{table*}

\begin{figure*}[h]
    \centering    \includegraphics[width=0.9\linewidth]{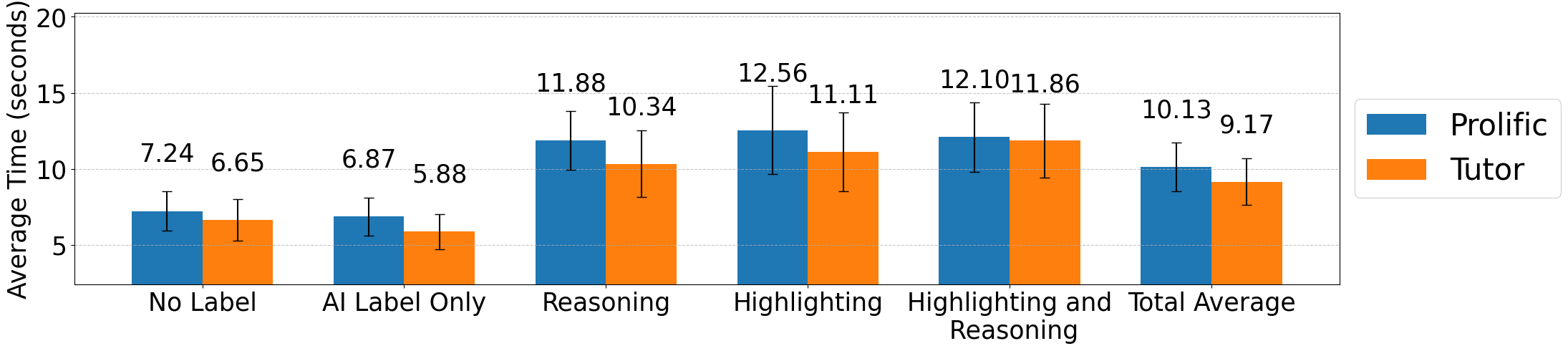}
    \caption{Average Time Spent by Condition and Group. The data is obtained after excluding outliers using the 1.5 interquartile range (IQR) method. Participants spend more time on conditions that have AI explanations. The error bar is Standard Error.}
    \label{fig:time_compare}
\end{figure*}

When further comparing performance between the two groups across conditions, as shown in \autoref{tab:performance_result}, we found that under the conditions of \textbf{Inline Highlighting, Highlighting + Reasoning}, and for \textbf{overall average}, the Prolific group significantly outperformed the Tutor group. The Prolific group also marginally significantly outperformed the Tutor group in the AI Label Only case. However, under the conditions \textbf{without any AI suggestions} and with \textbf{only Text Reasoning} explanations, the two groups had no significant difference.

\subsection{Time Spent comparison}

The results are presented in \autoref{fig:time_compare} and \autoref{tab:avg_time}. We conducted a repeated-measure 2-way ANOVA to examine the main effects of \textbf{Inline Highlighting} and \textbf{Text Reasoning}, as well as their interaction with the data after IQR filtering. The analysis revealed a significant main effect respectively for \textbf{Inline Highlighting} (\( F(1, 57) = 25.74, p < .001, \eta_p^2 = .311 \)) and \textbf{Text Reasoning} (\( F(1, 57) = 19.12, p < .001, \eta_p^2 = .251 \)). Additionally, there was a \textbf{significant interaction effect} between Highlighting and Reasoning (\( F(1, 57) = 18.26, p = .001, \eta_p^2 = .243 \)). This interaction suggests that the combined effect of \textbf{Inline Highlighting} and \textbf{Text Reasoning} does not lead to a substantial increase in time; rather, their effects partially offset each other.
Moreover, we found no significant difference in time spent between participants from Tutor and Prolific.


\subsection{Over-reliance}


\begin{table*}[h]
    \centering
    \caption{Over-Reliance T-test Comparison Results Between Tutor and Prolific Participants.}
    \begin{tabular}{c c c c l c}
        \hline
        Condition & Tutor Mean (SD) & Prolific Mean (SD) & t-statistic & p-value \\
        \hline
        Inline Highlighting & 0.654 (0.279) & 0.792 (0.215) & -2.696 & 0.0083** \\
        Inline Highlighting + Text Reasoning & 0.718 (0.253) & 0.839 (0.175) & -2.702 & 0.0082** \\
        AI Label Only & 0.660 (0.211) & 0.755 (0.222) & -2.151 & 0.0340* \\
        No Label & 0.516 (0.230) & 0.536 (0.225) & -0.440 & 0.6611 \\
        Text Reasoning & 0.750 (0.233) & 0.797 (0.245) & -0.954 & 0.3426 \\
        Overall Average & 0.660 (0.165) & 0.744 (0.114) & -2.896 & 0.0047** \\
        \hline
        \\[-1.8ex]
        {$\textsuperscript{*}p < 0.05$, $\textsuperscript{**}p < 0.01$, $\textsuperscript{***}p < 0.001$}
    \end{tabular}
    \label{tab:over_reliance_result}
\end{table*}

\begin{figure*}[h]
    \centering
    \includegraphics[width=0.9\linewidth]{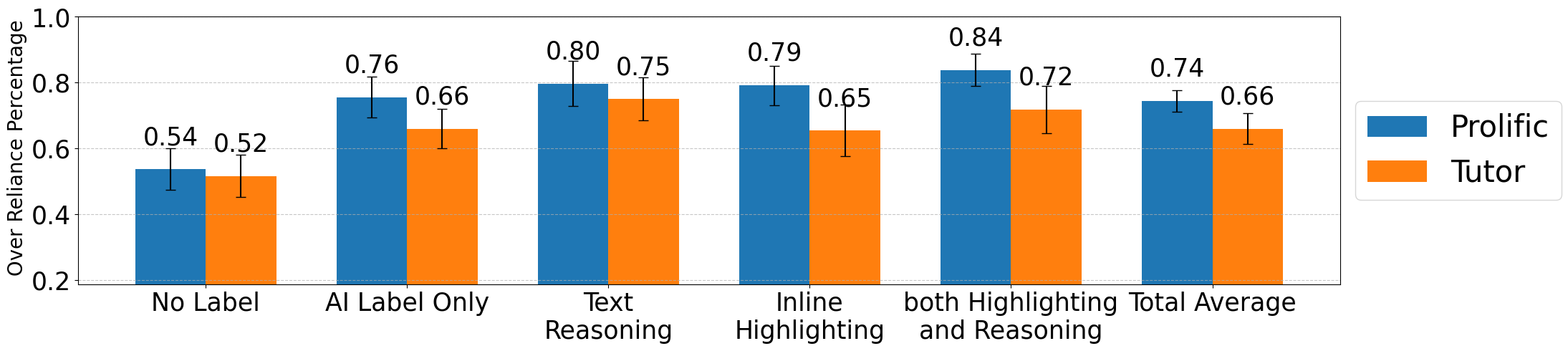}
    \caption{Average over reliance percentage (users were wrong when the AI was wrong) across conditions and groups. Participants from Prolific show higher over reliance. Error bars represent standard error.}
    \label{fig:over-reliance}
\end{figure*}

The results comparing over-reliance percentages for Tutors and Prolific participants across different task conditions are presented in \autoref{fig:over-reliance} and \autoref{tab:over_reliance_result}. We conducted a one-way ANOVA to investigate differences in over-reliance rates across conditions, revealing a highly significant effect \((F(5, 570) = 16.4074, p < .0001)\). Post-hoc analysis using Tukey’s HSD showed that, for questions where the AI was giving incorrect suggestions, participants who received AI suggestions showed significantly more errors compared to the baseline condition (all \(p_{\text{adj}} < .05\)). These findings indicate that the presence of wrong AI suggestions led participants to make wrong decisions, whether in the form of labels, \textbf{Text Reasoning}, or \textbf{Inline Highlighting}.

Additionally, the repeated-measures comparison indicated a significant main effect of \textbf{Text Reasoning} on over-reliance, (\( F(1, 94) = 8.81, p = .004, \eta_p^2 = .086 \)), whereas \textbf{Inline Highlighting} did not exhibit a significant effect, (\( F(1, 94) = 0.49, p = .482, \eta_p^2 = .005 \)). Furthermore, there was no significant interaction effect between \textbf{Inline Highlighting} and \textbf{Text Reasoning} (\( F(1, 94) = 0.069, p = .793, \eta_p^2 = .001 \)). These results suggest that the presence of \textbf{Text Reasoning} significantly increases participants' tendency toward overreliance, while \textbf{Inline Highlighting} does not have a notable impact.

When further comparing performance between the two groups across conditions, as shown in \autoref{tab:over_reliance_result} and \autoref{fig:over-reliance}, we found that under the conditions of \textbf{Highlighting, Highlighting + Reasoning}, \textbf{AI Label Only}, and for \textbf{overall average}, Prolific participants significantly exhibited more over-reliance compared to the Tutor. However, under \textbf{baseline} and \textbf{Text Reasoning} conditions, the two groups showed no significant difference.

\subsection{Under-reliance}

\begin{table*}[h]
    \centering
    \caption{Under-Reliance Comparison T-test Results Between Tutor and Prolific Participants.}
    \begin{tabular}{c c c c l c}
        \hline
        Condition & Tutor Mean (SD) & Prolific Mean (SD) & t-statistic & p-value \\
        \hline
        Inline Highlighting & 0.202 (0.155) & 0.099 (0.111) & 3.742 & 0.0003** \\
        Inline Highlighting + Text Reasoning & 0.160 (0.161) & 0.083 (0.112) & 2.721 & 0.0078** \\
        AI Label Only & 0.206 (0.168) & 0.134 (0.137) & 2.314 & 0.0228* \\
        No Label & 0.251 (0.151) & 0.223 (0.127) & 0.989 & 0.3254 \\
        Text Reasoning & 0.156 (0.163) & 0.115 (0.149) & 1.280 & 0.2038 \\
        Overall Average & 0.195 (0.122) & 0.131 (0.083) & 3.022 & 0.0032** \\
        \hline
        \\[-1.8ex]
        {$\textsuperscript{*}p < 0.05$, $\textsuperscript{**}p < 0.01$, $\textsuperscript{***}p < 0.001$}
    \end{tabular}
    \label{tab:under_reliance_result}
\end{table*}

\begin{figure*}[h]
    \centering
    \includegraphics[width=0.9\linewidth]{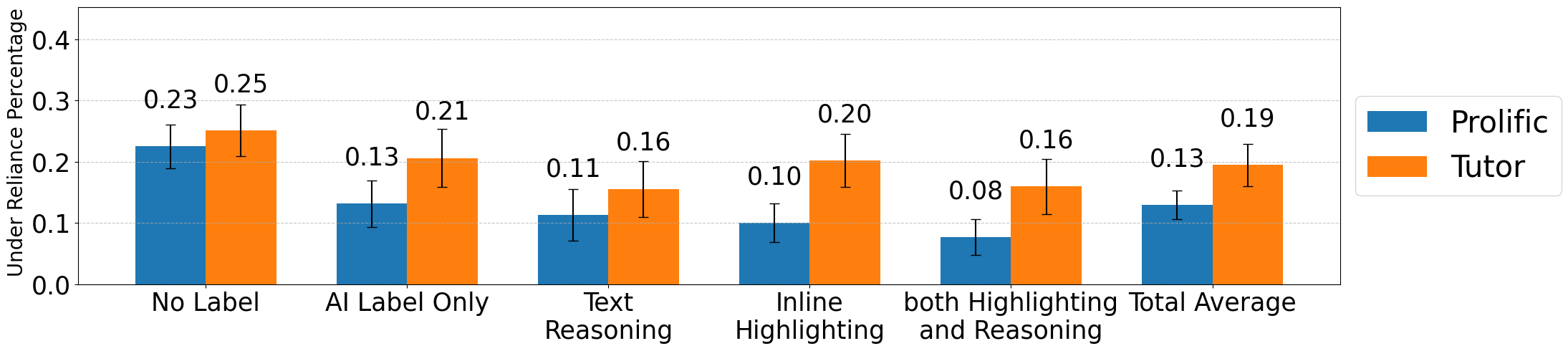}
    \caption{Average Under-Reliance Percentage, where users were wrong even if AI's suggestions are correct, across different Conditions and Groups. It can be observed that tutors are more prone to under-reliance. The error bar is Standard Error.}
    \label{fig:under-reliance}
\end{figure*}
The results comparing under-reliance percentages for Tutors and Prolific participants across different conditions are presented in \autoref{tab:under_reliance_result} and \autoref{fig:under-reliance}. A one-way ANOVA revealed a significant effect of condition on under-reliance (\(F(5, 570) = 7.9017, p < .001\)). Subsequent Tukey’s HSD post-hoc tests indicated that, for questions for which AI was correct, 
participants who received AI suggestions had significantly less error rate compared to the baseline condition (all \(p_{\text{adj}} = .05\)). This finding suggests that participants tend to trust the AI, and thus, when the AI is correct, they tend to respond correctly.

Additionally, the repeated-measures comparison indicated a significant main effect of \textbf{Text Reasoning} on underreliance (\( F(1, 94) = 8.71, p = .004, \eta_p^2 = .085 \)), whereas \textbf{Inline Highlighting} did not exhibit a significant effect (\( F(1, 94) = 2.28, p = .134, \eta_p^2 = .024 \)). Furthermore, there was no significant interaction effect between \textbf{Inline Highlighting} and \textbf{Text Reasoning} (\( F(1, 94) = 0.05, p = .944, \eta_p^2 = .000 \)). These results suggest that \textbf{Text Reasoning} effectively reduces underreliance, while \textbf{Inline Highlighting} does not have a notable impact.

When further comparing performance between the two groups across conditions, we found that under the conditions of \textbf{Inline Highlighting, Highlighting + Reasoning, and AI Label Only}, the Tutor group significantly exhibited more under-reliance compared to the Prolific group. However, under \textbf{baseline} and \textbf{Text Reasoning} conditions, the two groups had no significant difference.

\section{Discussion}


\subsection{Human-AI Collaboration vs. ``AI-Only'' Approaches.}
First, while participants do benefit from AI-generated explanations, particularly when both \textit{text reasoning} and \textit{highlighting} are provided, the improvement remains modest. Across most conditions, overall human-AI team accuracy still falls short of the AI-only baseline of 0.8. This suggests that even when collaborating with AI to judge the correctness of tutor responses, humans fail to achieve a complementary effect in decision-making tasks. In other words, unlike some earlier studies \cite{morrison2023impact,morrison2023evaluating,bansal2021does}, we did not observe a complementary effect in which experts working with AI outperformed either the AI alone or the human expert alone.
From a practical standpoint, these results pose a question about the necessity of human intervention in educational decision-making tasks. It may be the case that an ensemble of multiple AIs (e.g., a majority vote from different Large Language Models) surpasses both humans alone and human-AI teams. This trend raises concerns about efficiency and accountability: if an AI makes the final decision better than a human, and humans will be biased by AI, who is held accountable for errors? Future work should therefore explore reliable and ``responsible'' automation methods that ensure accountability, whether through transparency measures, oversight processes, or clear guidelines for error remediation.

\subsection{Text Reasoning, Highlighting, and Reliance Patterns Difference in Expertise}

Our findings are consistent with previous research showing that, despite its potential advantages, human judgment often leads to errors through either under-reliance or over-reliance on AI \cite{bansal2021does,vereschak2021evaluate,si2023large}. 
Moreover, we found that in the context of evaluating tutor responses, \emph{text reasoning} exerts the more pronounced influence on both sides of the trust dynamic. Specifically, \textit{textual reasoning} reduces \emph{under-reliance} (users become more convinced when AI is correct) but increases \emph{over-reliance} (users are less likely to reject incorrect AI suggestions). In contrast, \textit{inline highlighting} on its own did not show a significant effect. These findings align with previous research, which also shows that \textit{inline highlighting} can reduce reliance \cite{bansal2021does,si2023large,morrison2023impact,morrison2023evaluating}.
However, neither explanation strategy significantly improved performance nor reached a complementary effect, which differs from previous research~\cite{bansal2021does,vereschak2021evaluate,morrison2023impact,morrison2023evaluating}. This suggests that current approaches still have room for improvement. Future research in AI-assisted educational decision-making should thus explore more adaptive explanation styles and strategies that mitigate both over- and under-reliance while still leveraging the benefits of AI assistance.

Furthermore, our results indicate that \textbf{text reasoning} has a significantly stronger influence on reliance than \textbf{inline highlighting}. \textbf{Text reasoning} significantly reduced under-reliance but also increased over-reliance, showing that explanatory content can meaningfully shift user judgments and must therefore be designed with particular care. In contrast, \textbf{inline highlighting} had less measurable effect, suggesting a need for future work on more refined visual cues that guide attention without increasing bias.


We also observed notable differences between non-tutors recruited from Prolific (``\textbf{novices}'') and experienced tutors (``\textbf{experts}''). \textbf{Novices} showed a stronger tendency to trust AI suggestions, which increased their overall accuracy but also made them more vulnerable to over-reliance when the AI was incorrect. This pattern aligns with prior findings \cite{morrison2023impact}. \textbf{Experts}, by contrast, were more skeptical and sometimes overrode correct AI suggestions, leading to under-reliance errors. This pattern suggests that in educational decision-making contexts, domain expertise may increase confidence in one’s own judgment, even in cases where the AI is reliably correct and does better than a human alone.
This pattern differs from earlier studies in non-educational contexts \cite{morrison2023impact}.
Future efforts in designing human-AI collaboration systems in education should, therefore, account for users’ varying levels of expertise and self-efficacy, and adapt explanation strategies to balance trust and skepticism accordingly.

\subsection{Implications}
In addition to performance measures, indicators such as over-reliance, under-reliance, and time cost can serve as process-level indicators for modeling human-AI collaboration. These traces can be logged and analyzed to understand how explanation designs calibrate trust and efficiency across user groups. This framing positions human–AI decision-making as analyzable evidence for learning analytics, motivating future work on adaptive dashboards and monitoring tools for AI-assisted decision support. Additionally, these indicators can also potentially be used to help train humans to become better decision-makers in AI-assisted contexts.

\subsection{Limitations and Future Works}

Our study has two limitations; firstly, despite the potential advantages of AI assistance, our study revealed that participants actually spent \emph{more time} on AI-assisted tasks. The experimental context may encourage participants to try to be as accurate as possible. In real-world settings (e.g., busy teachers juggling many tasks late at night), decision-making could be more time-constrained or subject to different motivational factors. Further work is needed to investigate how time pressures influence human-AI interaction: for example, do educators just follow AI and skip carefully reading its explanations if they are overloaded in real-world decision-making scenarios?

Moreover, our study focused on a single type of decision-making task (evaluating the appropriateness of tutor praise). Other educational tasks, such as open-ended response evaluation or essay grading, may present different challenges for AI explanation and user reliance. Our findings, therefore, underscore the need for broader investigations into how AI-assisted decision-making applies across diverse educational contexts. Identifying consistent patterns of over-reliance and under-reliance can inform better interface design, explanation styles, and training protocols, ultimately fostering more effective collaboration between humans and AI in education.

\section{Conclusion}
Our findings highlight a complex reality: although AI explanations can improve performance under certain conditions, particularly for novices, designers must carefully manage the balance between over-reliance, under-reliance, and practical constraints such as time. By framing reliance and efficiency measures as learning analytics constructs, our study shows how AI-assisted decision-making can generate analyzable process data that informs both theory and practice. To fully harness AI’s potential in education, future research should explore adaptive explanation strategies grounded in learning analytics, investigate how these indicators can be embedded into real-time decision-support dashboards, and establish accountability frameworks that integrate human expertise with data-driven insights. In this way, human-AI collaboration can move beyond interface design to contribute directly to the broader learning analytics cycle of data collection, analysis, interpretation, and informed action.

\begin{acks}
This research was supported by the Generative AI + Education Tools R\&D Seed Grant at Carnegie Mellon University. We also thanks to the
National Science Foundation (award CNS-2213791) for partial support of this work.
\end{acks}

\bibliographystyle{ACM-Reference-Format}
\bibliography{reference}

\end{document}